\documentclass[%
reprint,
superscriptaddress,
longbibliography,
 amsmath,amssymb,
 aps,
prl,
]{revtex4-1}

\usepackage{graphicx}
\usepackage{dcolumn}
\usepackage{bm}
\usepackage{natbib}
\usepackage{color, soul}
\usepackage{overpic}
\usepackage{psfrag}
\usepackage{ulem}
\usepackage{siunitx}  
\usepackage{xcolor}

\definecolor{dablue}{rgb}{0,0.4,0.2}
\newcommand{\trr}[1]{\textcolor{black}{#1}}
\newcommand{\srr}[1]{} 
\newcommand{\tdd}[1]{\textcolor{black}{#1}} 
\newcommand{\sdd}[1]{}

\begin{document}

\title{\tdd{Dynamics of azimuthal thermoacoustic modes \\ in imperfectly symmetric annular geometries}}
\author{Thomas Indlekofer}
\email{thomas.indlekofer@ntnu.no}
\affiliation{Department of Energy and Process Engineering, Norwegian University of Science and Technology, Trondheim, Norway}
\author{Abel Faure-Beaulieu}
\affiliation{CAPS Laboratory, Department of Mechanical and Process Engineering, ETH Z\"urich}
\author{James R. Dawson}
\affiliation{Department of Energy and Process Engineering, Norwegian University of Science and Technology, Trondheim, Norway}
\author{Nicolas Noiray}
\email{noirayn@ethz.ch}
\affiliation{CAPS Laboratory, Department of Mechanical and Process Engineering, ETH Z\"urich}

\date{\today}

\begin{abstract}
Using \sdd{an} \tdd{a nominally symmetric}  annular combustor, we \trr{present} experimental evidence of \sdd{both explicit and}\tdd{a predicted} spontaneous \tdd{symmetry breaking} and \tdd{an unexpected} explicit symmetry breaking in the neighborhood of the Hopf bifurcation, which separates linearly-stable azimuthal thermoacoustic modes from self-\trr{oscillating modes}. We \trr{derive and solve a}\sdd{propose a model of Langevin equations, which can explain the observed predominantly standing modes with preferred nodal line direction in the vicinity of the bifurcation. We also present the solutions of the associated} \tdd{multidimensional} Fokker-Planck equation \sdd{to provide}\tdd{to unravel}  a \sdd{complete}\tdd{unified} picture of the phase space topology\sdd{, with transitions between unimodal, bimodal, single torus and double tori probability density functions of the eigenmode state vector}\tdd{. We demonstrate that symmetric probability density functions of the  thermoacoustic state vector are elusive, because the effect of \sdd{resistive and reactive} asymmetries, even imperceptible ones, is  magnified close\sdd{of the} to the bifurcation.} 

\end{abstract}

\pacs{Valid PACS appear here}

\maketitle
Symmetry breaking in systems where waves propagate along  \trr{closed loops or spheres} occur in a wide variety of physical problems. Examples range from Bose-Einstein condensates in toroidal traps \cite{Kolar2015,Marti2015}, whispering gallery modes in optomechanical resonators \trr{\cite{Shen2016657,Ceppe2019,Woodley2021}}, counter-propagating modes in \trr{optical ring cavities} \cite{Megyeri18}, nonreciprocal acoustic circulators \cite{Fleury2014516}, intrinsic rotation in Tokamak plasmas \cite{Rice2011},  to sporadic reversal in direction of planetary dynamo waves \cite{petrelis_2009,Sheyko2016551}. Symmetry breaking is also ubiquitous in fluid mechanics \cite{crawford91}, for example in von-K\'arm\'an swirling flows with counter-rotating toroidal vortices \trr{\cite{Burguete2007,Burguete2013,Faranda2017}}.  Recently, symmetry breaking was observed for azimuthal thermoacoustic modes in annular combustors typically found in jet engines or power generation gas turbines.  First described by Lord Rayleigh \cite{Rayleigh1878319}, thermoacoustic instabilities that arise from a constructive coupling between the acoustic field and the heat release rate of the flames lead to damaging vibrations.  Azimuthal thermoacoustic modes have been investigated in real engines \cite{krebs,seume,noi1}, laboratory experiments \cite{worth,worth3,bourgouin2} and numerical simulations \cite{wolf}. Over the last decade, the dynamic nature of these modes has received significant attention \cite{Bauerheim2016}, and the following classification has been established: standing modes, whose nodal line direction remains constant or slowly drifts around the annular chamber, spinning modes, whose nodal line spins at the speed of sound, or a combination of both referred to as mixed modes can exist. Several studies focused on the effects of \tdd{clearly identified} combustor asymmetries, e.g. \trr{\cite{worth2,bauerheim_jfm,bourgouin3,Berger2018,Kim2020}}, and several reduced-order models  \cite{polif,schuer,noi2,Ghirardo201652,Magri2016411,hummel18,moeck18,Orchini2019,Yang2019137,Buschmann2020251} were proposed to explain experimental observations with limited success. Notably, a recently proposed ansatz for describing the acoustic field \cite{quat} was combined with the wave equation to describe an ideal  annular combustor, unifying previous theoretical frameworks \cite{abel}. \sdd{However, a detailed analysis of the transition from linearly stable to self-sustained azimuthal thermoacoustic modes is missing. }\tdd{However, the transition from linearly stable to self-oscillating azimuthal thermoacoustic modes has never been investigated.} In this letter, we fill this gap with  experimental data\sdd{which shows} \trr{showing} that the route to a self-sustained quasi-pure spinning mode passes through stationary states that are characterized by a statistically prevailing standing mode and intermittent reversals of spinning direction. We also build upon the  model from \cite{abel} to unravel the complex topology of the phase space \sdd{and the underlying symmetry breaking. The modelling approach presented in this work is expected to have general relevance for systems in which waves propagate along toroidal and spherical geometries.}\tdd{ to demonstrate that symmetrically designed combustors will always display explicit symmetry breaking of the thermoacoustic modes at the bifurcation.}\\
\begin{figure}
    \centering\def\svgwidth{0.72\columnwidth}
    \input{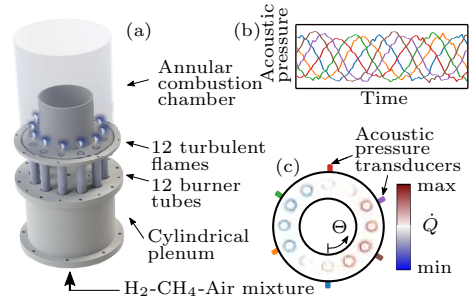}
\vspace{-3mm}
    \caption{(a) Model gas turbine combustor. (b) Acoustic pressure recorded with 6 transducers. During the selected interval, the thermoacoustic limit-cycle is a quasi-purely spinning eigenmode. (c) Phase-averaged flame chemiluminescence.}
    \label{fig:exp_setup}
\end{figure}{}
\noindent\textit{Experimental evidence of explicit and spontaneous symmetry breaking.}\textemdash\sdd{Experiments were conducted using the atmospheric annular combustion chamber described in \cite{worth2}  and sketched in Fig.~\ref{fig:exp_setup}. Briefly, a premixed fuel-air mixture is fed into a cylindrical plenum that conditions and divides the flow into twelve identical burners 
 \cite{supmaterial}. The fuel is composed of \SI{70}{\percent} hydrogen and \SI{30}{\percent} methane by power. The power is fixed to \SI{6}{\kilo\watt} per burner and the equivalence ratio $\Phi$ is varied in the range of 0.4875 to 0.575 with steps of 0.0125.}
\trr{Experiments were conducted using an annular combustor with 12 premixed flames operated at atmospheric condition (see  Fig.~\ref{fig:exp_setup}) and fueled with a mixture of $\SI{70}{\percent}$ H$_2$ and $\SI{30}{\percent}$  CH$_4$ by power. The thermal power was fixed to 72 kW and the equivalence ratio was varied between 0.4875 and 0.575.} Twelve 
 pressure transducers were mounted at six azimuthal positions flush with the inner wall of the burner tubes \cite{supmaterial} and recorded with a sampling rate of \SI{51.2}{\kilo\hertz}.  
  \begin{figure*}
    \centering\def\svgwidth{1\textwidth}
    \input{Images/intro_figure_rebuttal_mod.tex}
\vspace{-6mm}
    \caption{(a) SPL for increasing $\Phi$. (b) Selected intervals for $\Phi=0.55$, during which the azimuthal mode is mixed, with \trr{CW and CCW} spinning \sdd{component in the clockwise (CW) and counterclockwise (CCW)} directions. The angles in the legend correspond to the locations of the transducers. (c) Evolution of slow-flow variables that define the state of the azimuthal thermoacoustic mode, during the last 6 seconds of the record for $\Phi=0.55$, together with their PDFs. \tdd{(d) PDFs and time traces from the model with combined resistive and reactive asymmetries.}}
    \label{fig:intro}
\end{figure*}
 The natural chemiluminescence of the flames viewed from the exhaust is recorded with a high speed camera using the same optical system as in \cite{worth2}. For each $\Phi$, the combustor was ignited, run for at least 60~s to reach thermal equilibrium before recording 100~s of acoustic data while keeping the operating condition constant, and then turned off. 
 The sound pressure level (SPL) in the annular combustor is plotted in Fig.~\ref{fig:intro}(a) for four $\Phi$. When $\Phi=0.5$, the thermoacoustic system is linearly stable. The system's strongest resonance corresponds to an eigenmode whose azimuthal distribution  exhibits one acoustic wavelength along the annular chamber circumference, and whose frequency is about 1090 Hz. The resonance oscillation of the different modes is sustained by the inherent broadband acoustic forcing from the turbulent combustion process. For larger $\Phi$, the acoustic energy production resulting from the constructive thermoacoustic feedback exceeds the acoustic energy dissipation, and the dominant eigenmode becomes linearly unstable. Under these conditions, thermoacoustic limit cycles are reached at the aforementioned resonance frequency, e.g. \cite{noi3}. Their dynamics \srr{is}\trr{are} governed by the nonlinear coherent response of the flame to the acoustic field and by the turbulent fluctuations of the heat release rate. 
Figure \ref{fig:intro}(b) presents the acoustic pressure $p$ at three azimuthal locations during two short intervals (I) and (II) of the 100~s record for $\Phi=0.55$, \srr{and} for which the self-sustained azimuthal thermoacoustic mode spins in the counterclockwise (CCW) and in the clockwise (CW) direction. Recently, a new ansatz for the acoustic pressure field based on quaternions was proposed to describe azimuthal eigenmodes in annular combustors  \cite{quat}. A set of \trr{four}\sdd{4} state variables $A, \theta, \chi$ and $\varphi$, which vary slowly with respect to the fast acoustic time scale $2\pi/\omega$, and which can be extracted from the acoustic pressure time series, allows the following well-defined description of the state of an azimuthal thermoacoustic mode
\begin{equation}
   \tilde{p}\left(\Theta,t\right)=A\left(t\right)\:e^{i\left(\theta\left(t\right)-\Theta\right)}\:e^{-k\chi\left(t\right)}\:e^{j\left(\omega{t}+\varphi\left(t\right)\right)},
\label{eq1}
\end{equation}
where $t$ and $\Theta$ are the time and the azimuthal coordinate, and $i$, $j$ and $k$ are the quaternion units. The acoustic pressure is $\text{Re}(\tilde{p})$,
and $A$ is the real-valued positive slowly-varying amplitude.
The slowly varying angle $\theta$ describes the position of the anti-nodal line \cite{thetathetapluspi}. The slowly varying angle $\chi$ indicates whether the azimuthal eigenmode is a standing wave ($\chi=0$), a pure CW or CCW spinning wave ($\chi=\mp\pi/4$) or a mix of both for  $0<|\chi|<\pi/4$. The angle $\varphi$ stands for slow temporal phase drift. These variables fully describe the 
acoustic fields $p$ 
as a function of the time and azimuthal position.
Figure \ref{fig:intro}(c) shows the time traces of the extracted slow-flow variables together with their probability density functions (PDFs) for the last 6~s of the record, an interval that is representative of the stationary 
dynamics when $\Phi=0.55$.  The joint PDF of the first three state variables deduced from the 100~s record at $\Phi=0.55$ is also shown in Fig. \ref{fig:intro_bloch}(b), where $A$, $\theta$ and $2\chi$ are taken as spherical coordinates as shown in Fig.~\ref{fig:intro_bloch}(a).  
At the condition  $\Phi=0.55$, the amplitude of the azimuthal mode has a relatively small standard deviation compared to its mean value. The intermittent changes of sign of the nature angle $\chi$ correspond to changes of spinning direction of the mixed mode, as shown in Fig.~\ref{fig:intro}(b). 
\trr{Although}\sdd{Albeit} the difference is \sdd{fairly} small, the most probable orientation of the anti-nodal line of the mixed-mode's standing component is not the same whether the mode spins CW or CCW\tdd{, a phenomenon that had never been reported and that will be explained later}. 
\begin{figure}
    \centering\def\svgwidth{0.70\columnwidth}
    \input{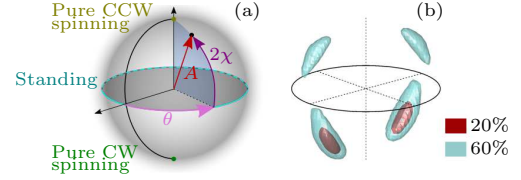}
    \vspace{-2mm}
    \caption{(a) Bloch sphere representation of the state of the azimuthal mode. (b) Experimental joint PDF for $\Phi=0.55$.}
    \label{fig:intro_bloch}
\end{figure}
\sdd{It is now worth examining}\trr{We now consider} the evolution of the state variable PDFs in the vicinity of the Hopf bifurcation. These stationary PDFs are shown in the left column of Fig.~\ref{fig:pdf}  for several 
$\Phi$. The evolution of $P(A)$, shown in Fig.~\ref{fig:pdf}(a), is typical of a supercritical Hopf bifurcation in presence of additive random forcing \cite{juel97}. The bifurcation occurs around $\Phi=0.5125$. This point separates  broadband-noise-driven resonances, when the azimuthal thermoacoustic mode is linearly stable, from limit cycles, when it is linearly unstable. Next, one can consider how $P(\varphi)$ changes 
with  $\Phi$ in Fig. \ref{fig:pdf}(d). Overall, the slowly varying temporal phase $\varphi$ drifts sufficiently during these 100 s records to yield rather flat distribution between 0 and $2\pi$. One can notice that some phases are statistically prevailing for the higher range of investigated $\Phi$, but these PDF maxima differ from case to case, and moreover, they wane when longer time traces are considered for computing the statistics, in agreement with the model which we propose in this letter. 
\begin{figure}	
\includegraphics[width=0.7\columnwidth]{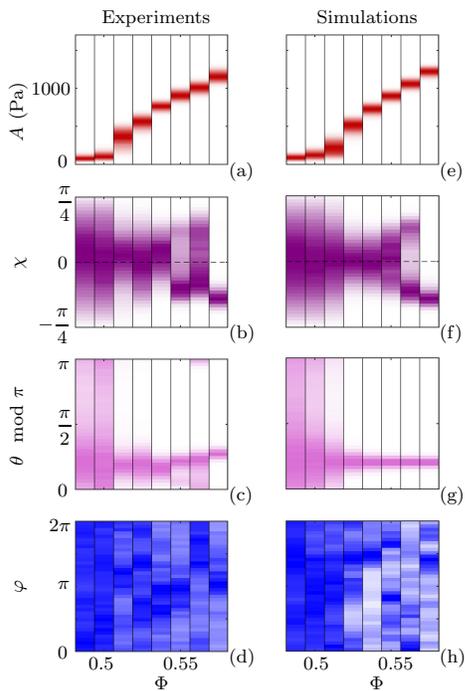}
	\vspace{-3mm}
	\caption{Stationary PDFs of $A$, $\chi$, $\theta$ and $\varphi$ at several $\Phi$ in the vicinity of the supercritical Hopf bifurcation. The PDFs are deduced from the slow-flow variable time traces, which are extracted from the acoustic pressure time traces  following the procedure in \cite{quat}. For (a) to (d), the acoustic time traces were obtained experimentally. For (e) to (h), the acoustic time traces were obtained by simulating 100 s of Eq.~\eqref{eq:wave}. Linear dependencies with respect to $\Phi$ of the parameters ($\nu$, $c_2\beta$, $\kappa$, $\Gamma$) were deduced from the data processing.}
	\label{fig:pdf}
\end{figure}
More interesting are the statistics of $\theta$ and $\chi$. Let us begin with $P(\theta)$ in   Fig. \ref{fig:pdf}(c): Even though the combustor was designed as rotationally symmetric as possible with a discrete rotational symmetry of order 12 (due to the 12 burners), the experimental results show a clearly preferred and repeatable anti-nodal line position for the different $\Phi$. 
Given the long run times and the high-order discrete rotational symmetry of the system,  one would either expect a quasi-uniform distribution of $\theta$ or a new preferred nodal line direction each time the combustor is ignited. The only explanation for this preferred orientation is the presence of a 
rotational symmetry breaking in the thermoacoustic system\sdd{(e.g. non-axisymmetric mixture injection at the bottom of the plenum)}. 
Furthermore, in addition to this explicit symmetry breaking, the system exhibits a spontaneous symmetry breaking in terms of the nature of the thermoacoustic mode. When $\Phi<0.55$, 
$P(\chi)$ is unimodal, with a maximum close to zero, indicating that the thermoacoustic mode is predominantly standing, which corroborates two recent theoretical studies \cite{abel,Ghirardo2021} that predicted the predominance of standing modes when the normalized stochastic forcing amplitude is large compared to the one of the limit cycle. For $\Phi=0.55$ and $\Phi=0.5625$,  $P(\chi)$  is bimodal: the spinning component of the mixed mode intermittently changes direction as illustrated in Fig.~\ref{fig:intro}(c). Now, when $\Phi=0.575$ in Fig.~\ref{fig:pdf}(b), these sporadic reversals of spinning direction do not take place anymore: the mixed mode spins in the CW direction and the symmetry of its state PDF is spontaneously broken beyond this threshold. It is worth noting that other runs performed for $\Phi\geq 0.575$, can also lead \trr{to} a robust CCW mixed mode. In the remainder of this letter, we present a \sdd{low-order theoretical} model \sdd{that captures both the explicit and spontaneous symmetry breaking} \tdd{to explain the unexpected explicit symmetry breaking}. 

\noindent\textit{Theoretical model and calibration of its parameters.}\textemdash In \cite{abel}, the quaternion  acoustic field 
was inserted into the following wave equation with distributed heat-release-rate fluctuations to model the 
thermoacoustic dynamics of an idealized thin annular chamber:
\begin{equation}\label{eq:wave}
\dfrac{\partial^2 p}{\partial t^2}  +  \alpha \dfrac{\partial p}{\partial t} - \dfrac{c^2}{\mathcal{R}^2} \dfrac{\partial^2 p}{\partial \Theta^2}
= (\gamma - 1)\:\dfrac{\partial \dot{Q}}{\partial t} + \Xi(\Theta,t). 
\end{equation}
In this equation, $\alpha$ is the acoustic damping at the annulus boundaries, $\mathcal{R}$ the annulus radius, $c$  the speed of sound, $\gamma$ the heat capacities ratio, $\dot{Q}(\Theta,t)$ the \textit{coherent} component of the heat release rate fluctuations and  $\Xi$ a spatially distributed white Gaussian noise  representing the effect of \textit{turbulent} fluctuations of the heat release rate \cite{Clavin199461,Lieuwen20033,Lieuwen200525,noi3}. The coherent heat release rate fluctuations are then  linked to the acoustic field using the  expression 
 $(\gamma-1)\dot{Q}=\beta\,[1+c_2\cos(2\Theta)] p -\kappa p^3$.
The first term stands for the linear response of the fluctuating heat release rate to acoustic perturbations. If $\beta>\alpha$ and $c_2=0$, the linear growth rate of the thermoacoustic mode $\nu=(\beta-\alpha)/2$ is positive, i.e. the mode is linearly unstable, and small acoustic perturbations are exponentially amplified. This pure exponential growth is not observed in real systems because of the inherent turbulence-induced stochastic forcing $\Xi$ and because of the nonlinearity of the flame response to the acoustic field beyond a certain level. The first term also accounts for spatial non-uniformities along the annulus, through the second component $c_2 \cos(2\Theta)$ of the Fourier expansion of any distribution of heat release rate source, which is the only 
component influencing the first azimuthal eigenmode \cite{noi1}. The second term, $-\kappa p^3$, defines the saturation of  flame response, which leads to limit cycle oscillations, and is sufficient to adequately capture the thermoacoustic dynamics in the vicinity of supercritical Hopf bifurcations  \cite{Lieuwen20033,noi4,LeeProci20}, as in the present experiment. Further discussion about this heat release rate model is given in \cite{supmaterial}. In \cite{abel}, spatial and temporal averaging are performed to derive a  system of coupled Langevin equations $\dot{\boldsymbol{Y}} = \boldsymbol{F}(\boldsymbol{Y}) + \boldsymbol{B}(\boldsymbol{Y}) \boldsymbol{N}$, governing the dynamics of the azimuthal mode state-vector  $\boldsymbol{Y}=(A,\chi,\theta,\varphi)^T$ via a deterministic contribution $\boldsymbol{F}(\boldsymbol{Y})$ and a stochastic forcing term  $\boldsymbol{B}(\boldsymbol{Y}) \boldsymbol{N}$, where $\boldsymbol{N}=(\zeta_A, \zeta_\chi, \zeta_\theta, \zeta_\varphi)^T$  is a vector of independent  white Gaussian noises of intensity $\Gamma/2\omega^2$. The latter intensity involves the constant $\Gamma$, which is the spatially-averaged stochastic forcing level originating from the turbulent fluctuations of the heat release rate. The expressions for $\boldsymbol{F}$ and $\boldsymbol{B}$ are given in \cite{abel,supmaterial} with the simplifying assumptions for their derivation. 
It is worth noting that this model cannot be used to predict the stability and slow-flow dynamics of a real combustor from the knowledge of its geometry and operating condition. However, we show here that 
its handful of parameters $[\nu; c_2\beta; \kappa; \Gamma]$ can be identified from the experimental data \trr{and then the model can reproduce the thermoacoustic dynamics}. Several methods are available for 
disentangling deterministic and stochastic contributions in stochastic trajectories, e.g. \cite{Frishman20,FerrettiPRX20,BrucknerPRL20}. Here, we used the method developed in \cite{Siegert_1998,Friedrich02,bottcher06,friedrich11}, which has already been taken up for investigating, among other areas, thermoacoustic and aeroacoustic problems, e.g. \cite{noi4,Boujo2020}. First, for each $\Phi$, the transition probabilities of the  state variables are extracted. Then, 
 these probabilities are used to compute the first and second Kramers-Moyal coefficients that serve as basis for parameter identification with least square fitting. More details about the robustness of the parameter identification \srr{is}\trr{are} given in \cite{supmaterial}. It was found that simple linear regression for the dependencies of $\nu$, $c_2\beta$, $\kappa$ and $\Gamma$ upon $\Phi$ allows \sdd{to capture}\trr{to unravel} the complex topology of the phase space. Indeed, time domain simulations of Eq.~\eqref{eq:wave}  were performed to generate \SI{100}{\second} acoustic time series for each $\Phi$, from which the state variables and their statistics were extracted. The results are presented in Fig.~\ref{fig:pdf}(e) to \ref{fig:pdf}(h), and are in close agreement with the experiments. This shows that despite the simplifications invoked to derive the low-order model, it can quantitatively reproduce the dynamics of the state variables.\\
\textit{Discussion}\textemdash The two coupled  equations for $A$ and $\chi$ 
recently allowed us to \trr{identify}\srr{sort out} the fundamental mechanisms involved in the spontaneous symmetry breaking of the thermoacoustic mode state. In fact, the symmetry of the phase space is not broken, but beyond a certain linear growth rate, the symmetry of the state may be, if the mode chooses either the CW or the CCW spinning state and the noise is not sufficiently large to allow sporadic escape from and return to this attractor \cite{abelproci}. Here, we go significantly further by considering the stationary solutions of the Fokker-Planck equation  governing the joint PDF of the state vector 
in order to provide a complete picture of the phase space topology (numerical method in  \cite{supmaterial}). 
\begin{figure}
  \centering\def\svgwidth{0.61\columnwidth}
  \input{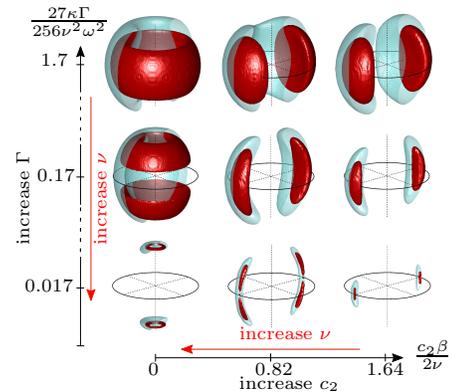}
    \vspace{-3mm}
    \caption{Stationary PDF of the eigenmode state computed with the Fokker-Planck equation for different noise intensity $\Gamma$ and asymmetry levels $c_2$ (state probability: 20\% and 60\%). The reference case in the center of the bottom row corresponds to the parameters of the calibrated model for $\Phi=0.55$.}
    \label{fig:final_balls}
\end{figure}
We consider the reference case $\Phi=0.55$, from which we arbitrarily change the linear growth rate $\nu$ and the level of asymmetry $c_2$ for fixed $\kappa$ and $\Gamma$. These stationary solutions are presented in Fig. \ref{fig:final_balls}. 
We start in the bottom right corner, where the state exhibits a \textit{unimodal} PDF corresponding to a statistically prevailing standing mode with fixed anti-nodal line orientation \cite{thetathetapluspi}. When the rotational asymmetry level is decreased, the PDF first becomes \textit{bimodal}, which corresponds to a predominant mixed  mode with intermittent changes of spinning direction. Ultimately, for $c_2=0$, the nodal line cannot lock anymore on a fixed orientation, it slowly drifts as a random walk and the PDF  takes the form of \textit{two tori} close to the poles. These results are in line with previous findings from \cite{noi1}, which gave an incomplete  description of the phase space topology. Then, the effect of noise and of the linear growth rate can be analyzed by moving vertically in the diagram. For an ideal rotationally symmetric combustor (first column) and for a fixed turbulent forcing level $\Gamma$ and saturation constant $\kappa$, the effect of decreasing the linear growth rate is seen by moving up: the \textit{two tori} merge when the system is brought close enough to the Hopf point (top row), which means that the system undergoes a transition from \trr{predominantly} spinning \trr{states} to  \trr{prevailing}  standing mode\trr{s} with \textit{single torus} PDF.   However, \tdd{a key consequence of the model, which had not been drawn so far, is that} this toroidal PDF will \sdd{presumably}not be observed in real combustors, because the more the source term gain $\beta$  tends to $\alpha$, and thus $\nu=(\beta-\alpha)/2\rightarrow 0$, the more $c_2\beta/(\beta-\alpha)$ will rise, which means that such decrease of $\beta$ would be accompanied with a move towards the right on the diagram. In other words, the effect of a \textit{resistive} asymmetry, however minute, will be magnified in the vicinity of the bifurcation
.\sdd{It should be noted that this magnifying lens effect would not occur in lossless systems. Future work will focus on incorporating the effect of \textit{reactive} asymmetry into the model, which may explain the observed small changes of $\theta$ accompanying the spinning direction changes. As a final remark, one can make the analogy between the Curie temperature above which permanent magnetic properties of certain materials are lost, and the present turbulence-induced stochastic forcing which, above a certain threshold, prevents robust spinning modes to exist.} \tdd{Furthermore,  we can add to the model tiny \textit{reactive} asymmetries \cite{reactive_as}. They also exist in real systems due to inherent geometrical imperfections, and cause, as shown in Fig. \ref{fig:intro}(d), the small changes of $\theta$ synchronized with changes of spinning direction and the predominance of CW spinning modes for $\Phi=0.55$. The spatial phase drift induced by these tiny imperfections is negligible during a few acoustic periods, but becomes significant over sufficiently long time scales. We can thus conclude that manufacturing annular systems free from explicit symmetry breaking at the bifurcation is out of reach. Possible direct applications of this framework include the analysis of non-radial stellar pulsations \cite{BUCHLER1995,buchler_1993,Aerts2021}, or the quest for weakly-dissipative nonreciprocal  devices based on an acoustic analog of the Zeeman effect \cite{Fleury2014516,Ding2019}, for which decreasing losses in circulators bring them closer to Hopf points where we show that slight asymmetries have considerable impact on wave dynamics.}

\begin{acknowledgments}
This project has received funding from the European Union's Horizon 2020 
program 
(Grant 
 765998). \end{acknowledgments}

\bibliography{references.bib}

\begin{thebibliography}{64}
\bibitem [{the()}]{thetathetapluspi}%
  \BibitemOpen
  \href@noop {} {\enquote {\bibinfo {title} {The eigenmode state is identical
  for $\theta$ and $\theta+\pi$.}}\ }\BibitemShut {NoStop}%
\bibitem [{rea()}]{reactive_as}%
  \BibitemOpen
  \href@noop {} {\enquote {\bibinfo {title} {In the model which combines
  resistive and reactive asymmetries, the last term of the l.h.s. of eq.
  \eqref{eq:wave} is multiplied by $[1+\varepsilon\cos(2[\theta-\vartheta])]$.
  fixing $\nu$, $\kappa$ and $\gamma$ to their values from fig. 4, the
  calibration for $\phi=0.55$ of this model variant yields $\varepsilon=0.0023$
  and $\vartheta=0.03$ and $c_2\beta=17$ s$^{-1}$.}}\ }\BibitemShut {NoStop}%
\end{thebibliography}






\end{document}